\documentclass[twocolumn,trackchanges]{aastex631}
\usepackage{CJK}
\usepackage{graphicx}
\usepackage{subfigure}
\usepackage{dsfont}
\usepackage{amsmath,amsfonts,amssymb,amsthm,extarrows}
\usepackage{xspace}
\usepackage{bm}
\usepackage{ragged2e}

\definecolor{myblue}{rgb}{0.25 0.35 0.95}
\hypersetup{linkcolor=blue,citecolor=blue,filecolor=blue,urlcolor=myblue}

\shorttitle{The Twisting of Radio Waves}
\shortauthors{Ze-Lin Zhang et al.}
\graphicspath{{./}{figures/}}

\begin{document}
\title{The Twisting of Radio Waves in a Randomly Inhomogeneous Plasma}

\correspondingauthor{Ze-Lin Zhang}
\email{zhangzl@ahstu.edu.cn; \\ryliu@nju.edu.cn}

\author[0000-0003-4621-0807]{Ze-Lin Zhang}

\affiliation{School of Electrical and Electronic Engineering, Anhui Science and Technology University, \\ Huangshan Avenue 1501, Bengbu 233030, China}
\author[0000-0003-1576-0961]{Ruo-Yu Liu}

\affiliation{School of Astronomy and Space Science, Nanjing University, Xianlin Road 163, Nanjing 210023, China}
\affiliation{Key laboratory of Modern Astronomy and Astrophysics, Nanjing University, Ministry of Education, Nanjing 210023, China}

\begin{abstract}
  Polarization of electromagnetic waves carries a large amount of information about their astrophysical emitters and the media they passed through, and hence is crucial in various aspects of astronomy. Here we demonstrate an important but long-overlooked depolarization mechanism in astrophysics: when the polarization vector of light travels along a non-planar curve, it experiences an additional rotation, in particular for radio waves. The process leads to depolarization, which we call `geometric' depolarization (GDP). We give a concise theoretical analysis of the GDP effect on the transport of radio waves in a randomly inhomogeneous plasma under the geometrical optics approximation. In the case of isotropic scattering in the coronal plasma, we show that the GDP of the angle-of-arrival of the linearly polarized radio waves propagating through the turbulent plasma cannot be ignored. The GDP effect of linearly polarized radio waves can be generalized to astrophysical phenomena, such as fast radio bursts and stellar radio bursts, etc. Our findings may have a profound impact on the analysis of astrophysical depolarization phenomena.
\end{abstract}


\section{Introduction} \label{sec:introduction}

According to Fermat's principle, light travels in straight lines when it propagates through an isotropic and homogeneous medium, the polarization state of it will remain constant over the whole path \citep{1960Landau}. Furthermore, while the medium is non-uniform, the polarization state might produce a nontrivial rotation to the polarization plane (also called Rytov rotation \citep{1938Rytov}, this is a classical example of the well-known geometric phase, or Berry phase \citep{1984Berry}, see \citealt{1992Anandan} and references therein) while the trajectory of the ray is not a planar curve, in other words, the torsion of the space curve is non-zero. Therefore, the rotation of the plane of polarization might give appreciable cumulative effects when the light travels a relatively long distance.

Half a century ago, under the geometrical optics approximation, Rytov rotation has already been discussed in the context of depolarization of linearly polarized light as it propagates through the Earth's atmospheric turbulence \citep{1970Kravtsov}. Together with the method calculating the depolarization of the linearly polarized light based on diffraction theory \citep{1967Strohbehn,1967Tatarskii}, both methods have demonstrated that the depolarization effects---`geometric' depolarization (GDP) and `diffractional' depolarization (DDP)---can be neglected for optical and microwave systems in the Earth's atmosphere \citep{2006Wheelon}. However, the results mentioned above may not hold outside the Earth's atmosphere for other wavelengths. For instance, environments like the solar radio flares via plasma emission mechanics \citep{2008Pick} and even in the burst of radio waves from magnetosphere of pulsars \citep{2022Beloborodov,2024Bacchini}.

Knowing the polarization behavior of radio emission is the key to understanding the radiation mechanisms of corresponding astrophysical processes and gaining insight into their turbulent plasma environments \citep{2022Lyutikov,2024Lower}. Now let us consider the propagation of linearly polarized radio waves in nonuniform turbulent plasmas. Due to the inhomogeneity of the medium, the propagation directions of radio waves vary from place to place and the Rytov polarization plane rotation occurs accordingly. If the medium is anisotropic, it can be inferred that this rotation will also contribute to the Faraday rotation angle as well \citep{2021Ferriere} and it will further refresh our understanding of the strength and geometry of astrophysical magnetic fields \citep{2010Broderick}. However, tracking the path of light in complex astrophysical environments is not an easy task. Thanks to the techniques of the numerical three-dimensional ray-tracing that are continually evolving \citep{2018MacDonald,2019Kontar}, we can simulate the paths of the rays in turbulent plasmas and analyze the depolarization processes via their torsion information.

There are various instabilities present in astrophysical plasmas. The analysis of instabilities is fundamental to the study of coherent radiation. The energy transfer in coherent radiation within the plasma can be achieved through the resonant wave-particle interaction \citep{2017Melrose}. In the case of weak magnetic field, non-thermal particles can excite Langmuir wave via beam-plasma instability in the plasma (the electromagnetic radiations are generated in anisotropic turbulent plasmas)~\citep{2024Bacchini}. The energy from Langmuir turbulence is partially converted into the energy of transverse radio waves and it primarily occurs near the fundamental frequency and the second harmonic of the plasma oscillation frequency. In the study of solar atmospheric observations, it has been found that many radio burst processes exhibit characteristics of coherent radiation, including that the frequency of radiation often appears near the plasma fundamental frequency or its second harmonic \citep{2017Kontar}.

In this paper, we consider the case that linearly polarized radio wave propagation from solar radio emission to Earth-based observers. \citet{2023Kansabanik} reported a robust imaging-based evidence for linearly polarized emission in metre-wavelength solar radio bursts. Due to the fact that the propagation path of light in a randomly inhomogeneous plasma is wavelength-dependent \citep{1991Tribble,2016Zhang}, and considering that the validity of the geometrical optics approximation is also wavelength-dependent \citep{1960Landau,1985Budden}, we assume that the radio waves generated in the outer solar corona during flares are created through the mechanisms of plasma emission for simplicity, then the radiation is generated close to the plasma frequency or its harmonic \citep{2008Pick}. Thus, our results become independent of the wavelength in the frequency range of 0.1--500 MHz \citep{2019Kontar}. We applied the three-dimensional stochastic description of ray and the techniques of differential geometry for the calculations \citep{1967Chernov,2019Kontar}. In the case of isotropic coronal scattering, the GDP of the angle-of-arrival of the radio waves propagating through the turbulent plasma is of the order of $1-10^3$ rad/AU by selecting appropriate parameters. Our results can be generalized to other astrophysical radio burst phenomena that exhibit coherent radiation characteristics (including plasma emission processes), such as radio emission from pulsar magnetospheres triggered by plasma instabilities.

\section{Theoretical Model} \label{sec:Model}

When investigating the propagation of radio waves in a turbulent medium, the solution based on geometrical optics approximation leads to the main constraint: $\lambda\ll l_i$, where $\lambda$ is the wavelength, $l_i$ is the inner scale of turbulent plasma. It means the wavelength is much smaller than the space scale of the inhomogeneity of the background field. Hence, the stochastic description of rays in three-dimensional space and the definition of the ray diffusion coefficient can be analyzed via differential geometry methods from a pedagogical viewpoint.

\begin{figure}[ht!]
  \centering
  \includegraphics[width=1.0\linewidth]{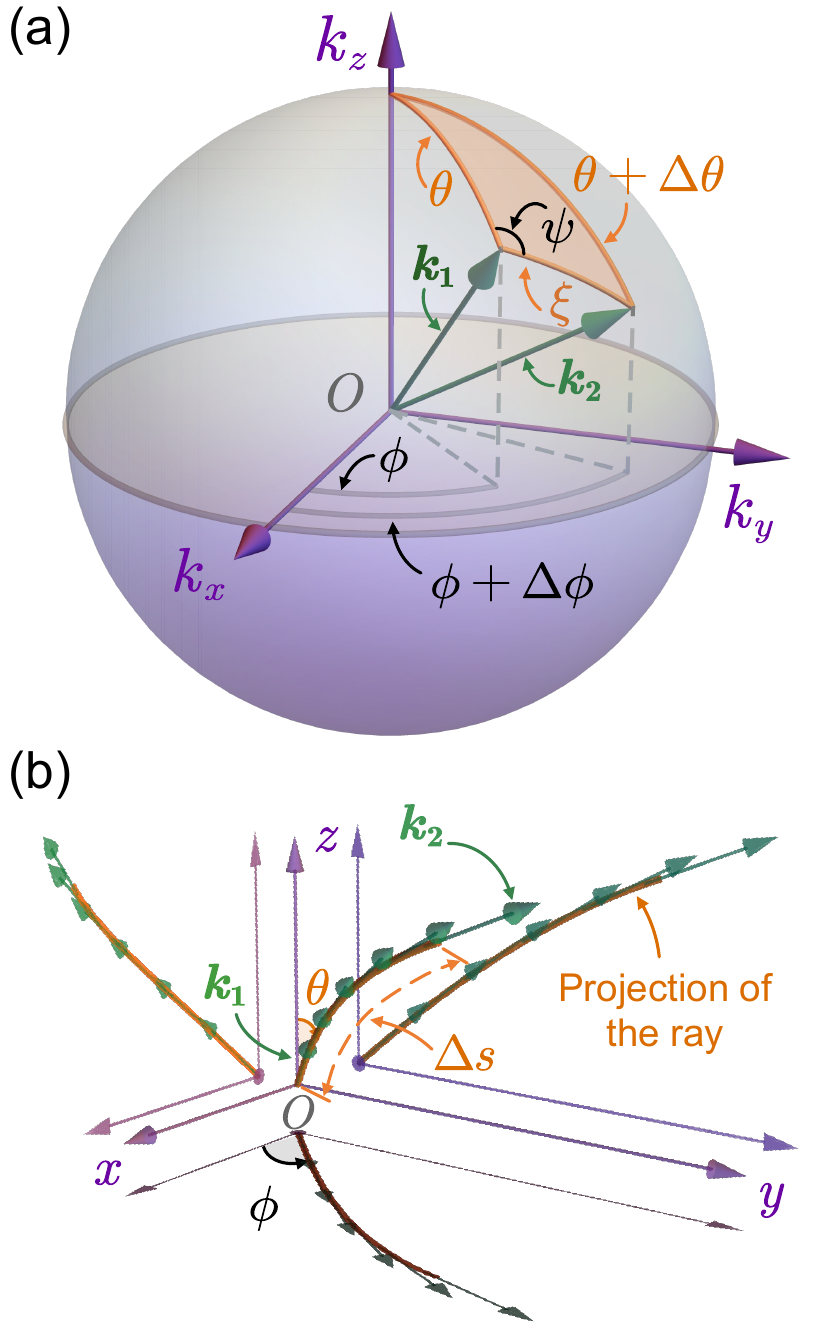}
  \caption{(a) The unit tangent vectors ($\textbf{\textit{k}}_1$ and $\textbf{\textit{k}}_2$) of the ray at different moments in unit $\textbf{\textit{k}}$-space (the tantrix sphere of the ray, or the sphere of tangent directions-$\textbf{\textit{k}}$). (b) The ray in $\mathbb{R}^3$ space and its projections onto the three mutually perpendicular coordinate planes ($x$-$O$-$y$, $y$-$O$-$z$ and $z$-$O$-$x$) are drawn by orange curves. $\Delta s$ is the arc-length between $\textbf{\textit{k}}_1$ and $\textbf{\textit{k}}_2$. The green arrows represent the unit wave vectors of light at different moments. The polar angle $\theta$ and azimuthal angle $\phi$ in the two subplots correspond to each other.
  \label{fig:Figure1}}
\end{figure}

\subsection{Three-dimensional Ray Statistics Model} \label{sec:Statistics}

\cite{1967Chernov} assumed a classic Gaussian spatial correlation function for the refractive index of a randomly inhomogeneous medium. The geometric description (known as the tantrix sphere, referring to \citealt{2006Berger} and \citealt{2017Thorne}) of this model is presented in Figure \ref{fig:Figure1}.

Let $\xi$ be the angle between two tangents at different moments to the ray, $\textbf{\textit{k}}_1$ and $\textbf{\textit{k}}_2$, separated by a distance $\Delta s$. If $\Delta s$ equals a few times the correlation distance $l_c$, then $\langle\xi\rangle=0$ and
\begin{eqnarray}\label{xi2}
\langle\xi^2\rangle=4\mathcal{D}\Delta s,
\end{eqnarray}
where $\langle\cdot\rangle$ denotes the value of ensemble average and $\mathcal{D}$ is defined as the ray diffusion coefficient. As shown in Figure \ref{fig:Figure1}, for tiny values of $\xi$, $\Delta\theta$ and $\Delta\phi$, the incremental changes of the polar angle $\theta$ will be given by
\begin{eqnarray}\label{dtheta}
\Delta\theta&\simeq&-\xi\cos\psi
\end{eqnarray}
and the azimuthal angle $\phi$ will be given by
\begin{eqnarray}\label{dphi}
\Delta\phi&\simeq&\xi\sin\psi/\sin\theta,
\end{eqnarray}
where $\psi$ is the angle between the vector $\textbf{\textit{k}}_2-\textbf{\textit{k}}_1$ and the plane spanned by $\textbf{\textit{k}}_2$ and the $z$ axis. The angles $\xi$, $\psi$, $\theta$, and $\phi$ are statistically independent of one another and of any angle in a different increment along the ray. The angle $\psi$ with uniform probability-density distribution satisfying $\langle\cos^2\psi\rangle=1/2$, together with Equations (\ref{xi2}) and (\ref{dtheta}) give
\begin{eqnarray}\label{dtheta2}
\langle\Delta\theta^2\rangle=2\mathcal{D}\Delta s.
\end{eqnarray}
The mean-square value of polar angle $\theta$ (known as the quivering angle) for the electromagnetic wave traveled a distance $s$ through the inhomogeneous medium will be given by
\begin{eqnarray}\label{theta2}
\langle\theta^2\rangle=2\mathcal{D} s.
\end{eqnarray}

\subsection{The Ray Diffusion Coefficient} \label{sec:Coefficient}

In the case of isotropic Gaussian spectrum of density fluctuations, the angular scattering rate per unit distance (also called the ray diffusion coefficient) is given by \citep{1967Chernov,1999Arzner,2019Kontar}
\begin{eqnarray}\label{epsilon2}
\mathcal{D}\equiv\mathcal{D}_{\theta\theta}=\frac{1}{2}\frac{\mathrm{d}\langle\theta^2\rangle}{\mathrm{d}s}
=\frac{\sqrt{\pi}}{4}\frac{\epsilon^2}{\sqrt{2}l_{c}}\frac{\omega_{pe}^4}{(\omega^2-\omega_{pe}^2)^2},
\end{eqnarray}
where $\epsilon^2={\langle\delta n^2\rangle}/{n^2}$ is the variance of relative density fluctuation and $n=\langle n\rangle$ is the ensemble average of plasma density, $l_c$ is the correlation length. The ray diffusion coefficient $\mathcal{D}_{\theta\theta}$ derived from the power-law spectrum of isotropic density fluctuations can be expressed in the same form by replacing $l_c$ with an equivalent scale length $l_{eq}$ given as \citep{2021Zhang}
\begin{eqnarray}\label{lc}
l_c\equiv l_{eq}=\pi^{-{3}/{2}}l_o^{2/3}l_i^{1/3},
\end{eqnarray}
where $l_i$ and $l_o$ are inner and outer scales delineating the inertial range of the turbulence separately.

\subsection{The Differential Geometry of Polarized Waves} \label{sec:Depolarization}

\begin{figure}[t!]
  \centering
  \includegraphics[width=0.8\linewidth]{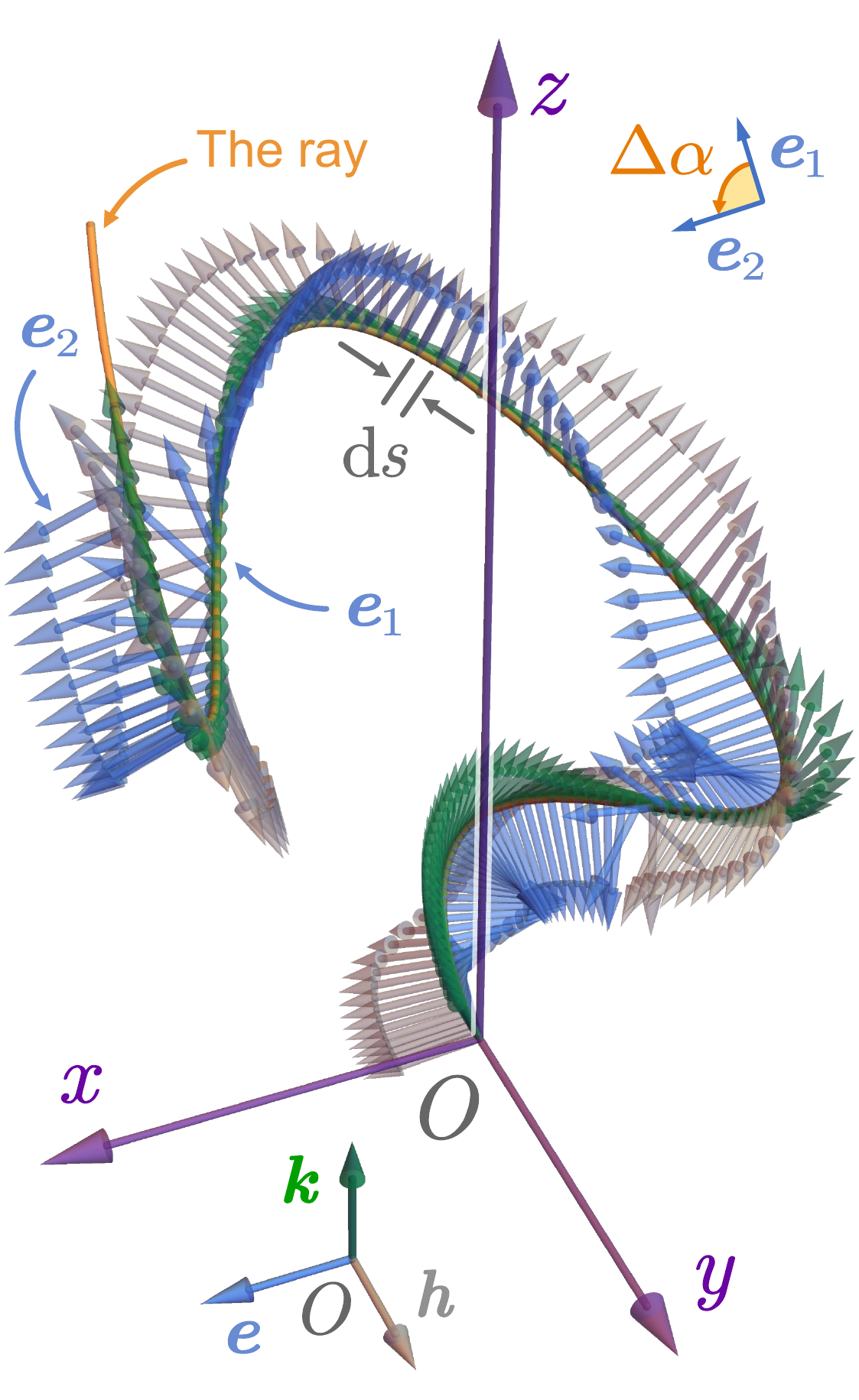}
  \caption{Polarization trihedral (represented by green, blue, and gray arrows, separately) transports along the ray (represented by orange curve) in $\mathbb{R}^3$ space with non-zero torsion. The angle $\Delta\alpha$ between the electric field vectors $\textbf{\textit{e}}_1$ and $\textbf{\textit{e}}_2$ at different moments is equal to the geometric phase.
  \label{fig:Figure2}}
\end{figure}

As shown in Figure \ref{fig:Figure2}, we define a polarization trihedral ($O$-$\textbf{\textit{k}}$-$\textbf{\textit{e}}$-$\textbf{\textit{h}}$, PT for short), where $\textbf{\textit{k}}$, $\textbf{\textit{e}}$ and $\textbf{\textit{h}}$ are the wave, electric-field and magnetic-field vectors of the ray at any infinite tiny increment $\mathrm{d}s$. The right-handed orthonormal frame $O$-$\textbf{\textit{k}}$-$\textbf{\textit{e}}$-$\textbf{\textit{h}}$ is called the Darboux frame \citep{1989Stoker}, which rotates an angle $\Delta\alpha$ relative to the Frenet-Serret frame ($O$-$\textbf{\textit{k}}$-$\textbf{\textit{n}}$-$\textbf{\textit{b}}$, where $\textbf{\textit{n}}$ and $\textbf{\textit{b}}$ are the normal and binormal vectors of the ray). Using theories of differential geometry \citep{2004Chruscinski}, we have
\begin{eqnarray}\label{knb1}
\textbf{\textit{k}}=\frac{\mathrm{d}\textbf{\textit{r}}}{\mathrm{d}s},\,\,\,\,
\textbf{\textit{n}}=\frac{1}{\kappa}\frac{\mathrm{d}^2\textbf{\textit{r}}}{\mathrm{d}s^2},\,\,\,\,
\textbf{\textit{b}}=\frac{1}{\kappa}\frac{\mathrm{d}\textbf{\textit{r}}}{\mathrm{d}s}\times\frac{\mathrm{d}^2\textbf{\textit{r}}}{\mathrm{d}s^2},
\end{eqnarray}
where $\kappa$ is the curvature of the ray $\textbf{\textit{r}}(s)$. Since the light waves are transverse, the vectors $\textbf{\textit{e}}$ and $\textbf{\textit{h}}$ are always co-planar with vectors $\textbf{\textit{n}}$ and $\textbf{\textit{b}}$. Then we have
\begin{eqnarray}\label{eh}
\begin{aligned}
\textbf{\textit{e}}&=(\cos\Delta\alpha)\textbf{\textit{n}}+(\sin\Delta\alpha)\textbf{\textit{b}},\\
\textbf{\textit{h}}&=-(\sin\Delta\alpha)\textbf{\textit{n}}+(\cos\Delta\alpha)\textbf{\textit{b}}
\end{aligned}
\end{eqnarray}
and the angle $\Delta\alpha$ is determined by the torsion $\tau$ of the spacial ray as $\Delta\alpha=-\tau \Delta s$, and it is also known as the geometric phase mentioned above (or topological phase, see \citealt{1990Vinitskii}), as presented in Figure \ref{fig:Figure2}. To complete the geometrical optics prescription of the field, a construction rule should be added: the PT does not rotate about the ray, which means
\begin{eqnarray}\label{opk}
\mathbf{\Omega}_p\cdot\textbf{\textit{k}}=0,
\end{eqnarray}
where $\mathbf{\Omega}_p$ is the angular velocity vector of the PT \citep{1966Lewis}. To examine $\mathbf{\Omega}_p$, let us identify arc-length $s$ with time $t$, and introduce the Frenet-Serret equations
\begin{eqnarray}\label{knb2}
\frac{\mathrm{d}\textbf{\textit{k}}}{\mathrm{d}t}=\kappa\textbf{\textit{n}},\,\,\,\,
\frac{\mathrm{d}\textbf{\textit{n}}}{\mathrm{d}t}=-\kappa\textbf{\textit{k}}+\tau\textbf{\textit{b}},\,\,\,\,
\frac{\mathrm{d}\textbf{\textit{b}}}{\mathrm{d}t}=-\tau\textbf{\textit{n}}.
\end{eqnarray}

If $\mathbf{\Omega}=\omega_1\textbf{\textit{u}}_1+\omega_2\textbf{\textit{u}}_2+\omega_3\textbf{\textit{u}}_3$ is the angular velocity of the moving trihedral $O$-$\textbf{\textit{u}}_1$-$\textbf{\textit{u}}_2$-$\textbf{\textit{u}}_3$ (a rigid body), then from the properties of  $\mathbf{\Omega}:\mathrm{d}\textbf{\textit{u}}/\mathrm{d}t=\mathbf{\Omega}\times\textbf{\textit{u}}$ and Equations (\ref{eh}) and (\ref{knb2}), we obtain $\textbf{\textit{u}}_1=\textbf{\textit{k}}$, $\textbf{\textit{u}}_2=\textbf{\textit{e}}$, $\textbf{\textit{u}}_3=\textbf{\textit{h}}$, $\omega_1=0$, $\omega_2=\kappa\sin\Delta\alpha$, $\omega_3=\kappa\cos\Delta\alpha$, then the angular velocity of PT is given by
\begin{eqnarray}\label{omegap1}
\mathbf{\Omega}_p=(\kappa\sin\Delta\alpha)\textbf{\textit{e}}+(\kappa\cos\Delta\alpha)\textbf{\textit{h}}=\kappa\textbf{\textit{b}}
\end{eqnarray}
which implies Equation (\ref{opk}). Compare Equation (\ref{omegap1}) with Equation (\ref{knb1}), we have
\begin{eqnarray}\label{omegap2}
\mathbf{\Omega}_p=\frac{\mathrm{d}\textbf{\textit{r}}}{\mathrm{d}s}\times\frac{\mathrm{d}^2\textbf{\textit{r}}}{\mathrm{d}s^2}.
\end{eqnarray}

Referring to \cite{1967Saleh}, the change of the polarization angle $\Delta\alpha$ can be defined as the cumulative rotation of the polarization vector around $z$ axis and then
\begin{eqnarray}\label{dalpha1}
\Delta\alpha=\mathbf{\Omega}_p\cdot\textbf{\textit{e}}_z\Delta s,
\end{eqnarray}
where $\textbf{\textit{e}}_z$ is the unit vector of the $z$ axis. Now we set
\begin{eqnarray}\label{drds}
\frac{\mathrm{d}\textbf{\textit{r}}}{\mathrm{d}s}=(\sin\theta\cos\phi,\,\,\sin\theta\sin\phi,\,\,\cos\theta).
\end{eqnarray}
Applying Equations (\ref{dphi}), (\ref{knb1}), (\ref{omegap1}), (\ref{dalpha1}-\ref{drds}) and noting that $\Delta\phi\simeq\mathrm{d}\phi/\mathrm{d}s\Delta s$, then we have
\begin{eqnarray}\label{dalpha2}
\Delta\alpha\simeq\sin^2\theta\Delta\phi=\xi\sin\theta\sin\psi,
\end{eqnarray}
we noticed that $\langle\xi\rangle=0$ implies $\langle\Delta\alpha\rangle=0$. From the Equations (\ref{xi2}) and (\ref{dtheta2}) and the fact that $\xi$, $\theta$ and $\psi$ are statistically independent, then gives
\begin{eqnarray}\label{dalpha21}
\langle(\Delta\alpha)^2\rangle=4\mathcal{D}^2 s\Delta s,
\end{eqnarray}
we noticed that $\langle(\Delta\alpha)^2\rangle$ depends on the position along the ray, $s$. Equation (\ref{opk}) indicates the polarization vector does not rotate around the ray itself, then we can conclude that $\Delta\alpha$ should be related to $\theta$, see Equation (\ref{theta2}). Summing the uncorrelated $\Delta\alpha$ of different increments along the ray of length $L$. The total change of the polarization angle as $\langle\alpha\rangle=0$ and $\langle\alpha^2\rangle=2\mathcal{D}^2L^2$. Compare with Equation (\ref{theta2}), we have $\sqrt{\langle\alpha^2\rangle}\sim\langle\theta^2\rangle$.

Substituting for the ray diffusion coefficient $\mathcal{D}$ from Equation (\ref{epsilon2}), we can estimate the root-mean-square (r.m.s) value of the polarization angle as
\begin{eqnarray}\label{rms}
\sqrt{\langle\alpha^2\rangle}&\equiv&\sqrt{2}\mathcal{D}_{\theta\theta}L
=\frac{\sqrt{\pi}}{4}\epsilon^2\frac{\omega_{pe}^4}{(\omega^2-\omega_{pe}^2)^2}\frac{L}{l_{c}}.
\end{eqnarray}
This is the key formula in this paper, and the following calculations are based on this formula to estimate the impact of GDP on the angle-of-arrival. It was found to increase linearly with the length of the path $L$.

\section{Results} \label{sec:Results}
\subsection{GDP in a Randomly Inhomogeneous Plasma} \label{sec:GDP}

\begin{figure*}[t!]
  \centering
  \includegraphics[width=1.0\linewidth]{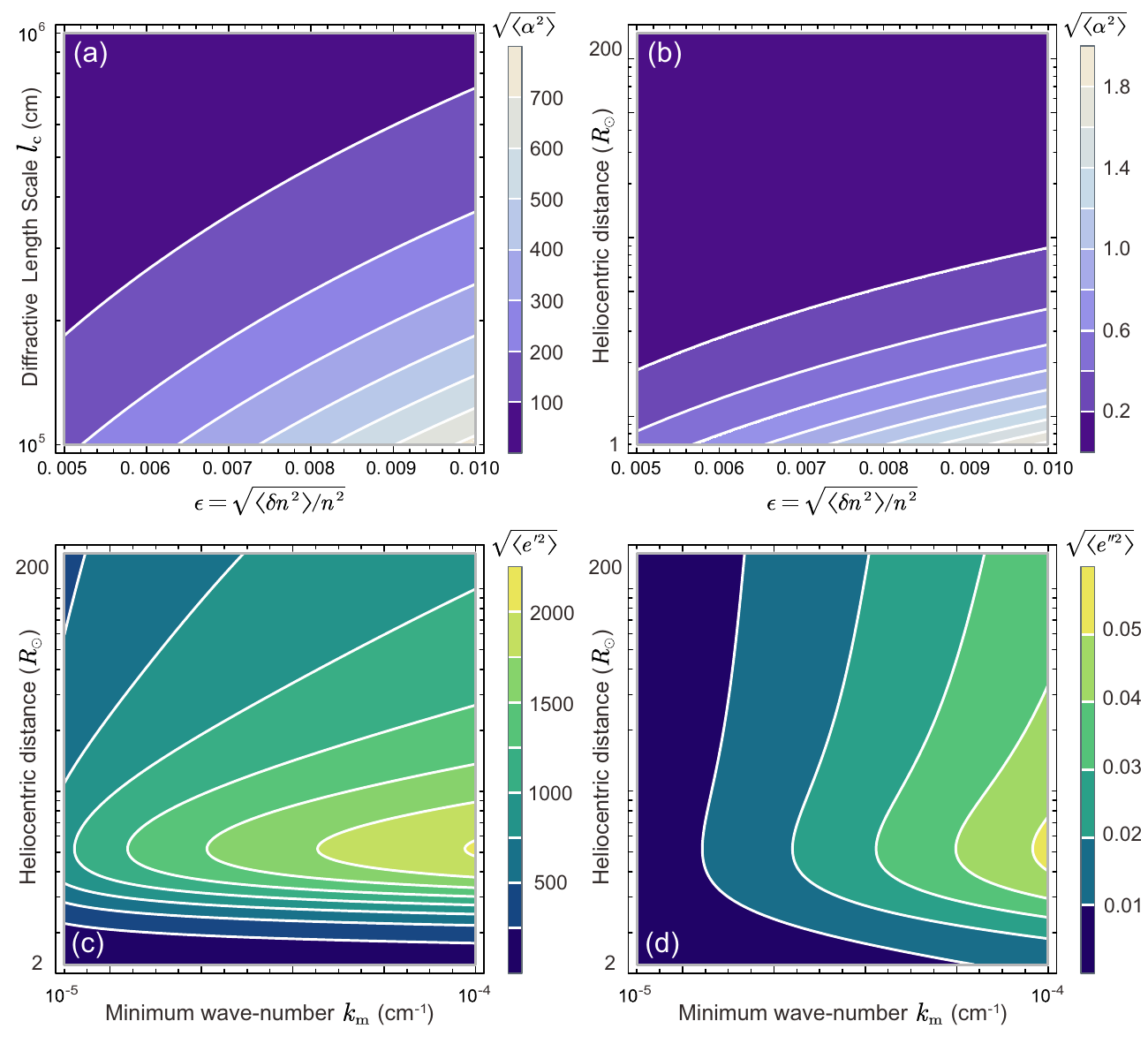}
  \caption{Upper panels: (a) The r.m.s. values of GDP $\sqrt{\langle\alpha^2\rangle}$ against $\epsilon$ and diffractive length scale $l_c$. (b) The r.m.s. values of GDP $\sqrt{\langle\alpha^2\rangle}$ against $\epsilon$ and the heliocentric distance $R_{\odot}$ which is from the Sun to the Earth. Lower Panels: (c) The r.m.s. values of quantity $\langle e^{\prime 2}\rangle$ against the heliocentric distance $R_{\odot}$ and the minimum wave-number $k_{\mathrm{m}}$ from the Sun to the Earth. (d) The r.m.s. values of quantity $\langle e^{\prime\prime 2}\rangle$ against the heliocentric distance $R_{\odot}$ and the minimum wave-number $k_{\mathrm{m}}$ from the Sun to the Earth.
  \label{fig:Figure3}}
\end{figure*}
Despite recent observations indicating that the scattering of radio waves in coronal plasma exhibits anisotropy~\citep{2017Kontar}, for simplicity, we continue to assume the isotropic scattering and it does not affect the discussion of the GDP effect. The main reasons for the above consideration are three: firstly, under the geometrical optics approximation, an isotropic and smoothly inhomogeneous medium is equivalent to an anisotropic medium~\citep{2004Bliokh}, it is sufficient to make the torsion of the ray path non-zero; secondly, in the case of anisotropic scattering, the impact of the background magnetic field on polarization should be considered, such as the Faraday depolarization process, while this will not impact the GDP process, it will complicate the issue; thirdly, plane waves propagated in an anisotropic medium are completely linearly polarized in certain polarization planes~\citep{1960Landau}, and it is consistent with the polarization state we considered here. Since the refractive index of the unmagnetized turbulent (isotropic and inhomogeneous) plasma $n_{\mathrm{ref}}=(1-\omega_{pe}^2/\omega^2)^{1/2}$ is significantly deviate from unity for the angular frequency of radio waves $\omega$ close to the local plasma frequency $\omega_{pe}$ in the turbulent plasma of the solar atmosphere, the impact of the density inhomogeneity along the wave path is significant in the transmission of solar radio bursts generated by plasma processes.

Assuming the radio point source with an isotropic distribution of directions $\textbf{\textit{k}}$ of the ray and with a frequency $\omega=2\omega_{pe}$ (second harmonic emission). We adopt the square root of the variance of relative density fluctuation $\epsilon$ varies from 0.05 to 0.1 \citep{2018Krupar}. The distance from source to observer $L=1$ AU $\simeq1.496\times10^{13}$ cm. For radio waves with frequencies from 3 MHz to 300 MHz, the Fresnel scale $r_{\mathrm{F}}=\sqrt{\lambda L/2\pi}$ varies from $1.54\times10^7$ cm to $1.54\times10^8$ cm, this is consistent with the description in \citealt{1992Narayan}. In the strong scattering environment appropriate for electromagnetic waves near the plasma frequency, the waves quickly become isotropic. In the case of strong scattering close to the sun, $r_{\mathrm{F}}$ should be much larger than the scale of the `diffractive' length $l_c$ (the correlation scale in Equation (\ref{epsilon2})), so we set it within the range of $10^{5}$ cm to $10^{6}$ cm. According to Equation (\ref{rms}), the r.m.s. values of GDP are close to $10^3$ rad/AU which is shown in Figure \ref{fig:Figure3} (a). The above calculations are performed in the case of the isotropic Gaussian spectrum of density fluctuations, the plasma density fluctuations can be considered as effectively static since the group velocity of density fluctuations is much less than the speed of light.

In addition, in situ observations indicate an inverse power-law spectrum of density fluctuations $\propto \textbf{\textit{q}}^{-(p+2)}$ with the exponent $p\approx5/3$ as observed and $\textbf{\textit{q}}$ is the wave-vector of electron density fluctuations~\citep{2013Alexandrova}. The inner scale of the solar wind turbulence $l_i=(r/R_{\odot})\times10^5~\mathrm{cm}$ can be viewed as a dissipation scale (the electron gyro-radius)~\citep{2013Alexandrova,2019Verscharen}, where the solar radius $R_{\odot}=6.955\times10^{10}$ cm, and the heliocentric distance $r$ varies from $R_{\odot}$ to $215 R_{\odot}$ (photon propagates from the Sun to the Earth). We choose the outer scale $l_o=0.25R_{\odot}(r/R_{\odot})^{0.82}$~\citep{2021Zhang}. According to Equation (\ref{lc}), $l_c$ varies from $5.6\times10^{7}$ cm to $6.3\times10^{9}$ cm where the geometrical optics approximation still holds for the radio waves with wavelengths from $10^2$ cm to $10^4$ cm. According to Equation (\ref{rms}), the maximum value of r.m.s. of is close to 2 rad/AU which is shown in Figure \ref{fig:Figure3} (b).

The three-dimensional Monte Carlo ray-tracing simulations by \citet{2020Chen} indirectly support our theoretical results, suggesting that space curve (the path of the ray) with torsion will cause linearly polarized radio waves to accumulate a geometric phase (the angle-of-arrival related to the GDP process) as it propagates through the isotropic and randomly inhomogeneous plasma. What needs to be noted is that if the polarization surface returns to its initial
direction upon reaching the observer, different gauges only bring
differences of integer multiples of $2\pi$, which has no corresponding observable effects.

\subsection{GDP vs. DDP} \label{sec:GDP vs. DDP}

\begin{figure}[t!]
  \centering
  \includegraphics[width=1.0\linewidth]{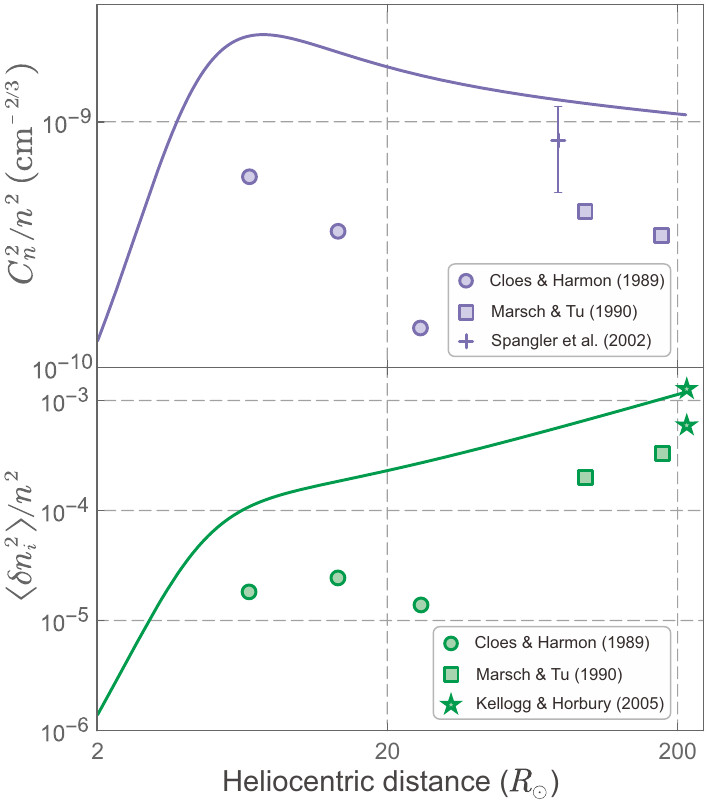}
  \caption{Top side: the variation of the normalized quantity ${C_{n}^{2}(r)}/{n^2(r)}$ with the heliocentric distance $R_{\odot}$ which varies from the Sun to the Earth. Bottom side: the spatial evolution trend of squared fractional density perturbation amplitude ${\langle\delta n_i^2\rangle}/{n^2}$ as it changes with the heliocentric distance $R_{\odot}$, where $l_i(r)=2.5\times10^4(r/R_{\odot}-1)^{1.3}~\mathrm{cm}$, see Equation (\ref{ni2}) and \citet{2023Kontar}. The data presented in the figure is derived from~\citealt{1989Coles,1990Marsch,2002Spangler,2005Kellogg}. The deviation between data and simulation is essentially controlled within an order of magnitude.
  \label{fig:Figure4}}
\end{figure}

To compare the effects of GDP and DDP on the propagation of radio waves in an inhomogeneous and isotropic plasma, we adopt the density profile
$n(r)~[\mathrm{cm}^{-3}]$ of the plasma used for the solar scattering simulations by \citet{2020Chen} and \cite{2019Kontar,2023Kontar} is
\begin{eqnarray}\label{nr}
n(r)=&&4.8\times10^9\left(\frac{R_{\odot}}{r}\right)^{14}+3\times10^8 \left(\frac{R_{\odot}}{r}\right)^6\cr\cr
&&+1.4\times10^6 \left(\frac{R_{\odot}}{r}\right)^{2.3},
\end{eqnarray}
where the solar radius $R_{\odot}=6.955\times10^{10}$ cm, and the heliocentric distance $r$ varies from $2 R_{\odot}$ to $215 R_{\odot}$. The amplitude of the \citet{1941Kolmogorov} density turbulence spectrum varies with distance $r$ from the Sun as
\begin{eqnarray}\label{cn2}
C_{n}^{2}(r)\simeq3.5\times10^3\left(\frac{r}{R_{\odot}}-1\right)^{-4.7} ~\textrm{cm}^{-20/3},
\end{eqnarray}
it is the normalization coefficient of the power spectrum and is directly related to the density variance, then from Equations (\ref{nr}) and (\ref{cn2}), we can calculate the normalized quantity ${C_{n}^{2}(r)}/{n^2(r)}~[\textrm{cm}^{-2/3}]$, as shown in the top side of Figure \ref{fig:Figure4} and it close to a constant for distances $>10 R_{\odot}$. The profiles of the dissipation length scale $l_i$ allow us to estimate the amplitude of the density fluctuations at the dominant inner scale. The squared fractional density perturbation amplitude at  $l_i$ is
\begin{eqnarray}\label{ni2}
\frac{\langle\delta n_i^2\rangle}{n^2}=4\pi l_i^{2/3}\frac{C_{n}^{2}(r)}{n^2(r)}.
\end{eqnarray}
It is close to the values of $\epsilon^2$ which are shown in the upper panels of Figure \ref{fig:Figure3} when the ray travels more than $10R_{\odot}$ away from the Sun, as shown in the bottom side of Figure \ref{fig:Figure4}. The low rotation measure in metre-wavelength solar radio bursts indicates that the linear polarized emission has encountered much lower electron densities, implying that it originates at a much higher position within the solar corona \citep{2024Dey}.

Again, assuming $L=1$ AU, the wavelength $\lambda=300$~cm, wave-number $k=2\pi/\lambda$. Following \citet{1970Kravtsov}'s train of thought, we estimate the quantities $\langle e^{\prime 2}\rangle$ (where $\textbf{\textit{e}}$ is the electric-field polarization vector as mentioned in Section \ref{sec:GDP}, its mean-square value is proportional to the square of the angle-of-arrival variance in Equation (\ref{rms})) and $\langle e^{\prime\prime 2}\rangle$ (as the average intensity of the orthogonal component of the field divided by that of the incident plane wave, see \citealt{2006Wheelon} for more details) with Kolmogorov's spectral density
\begin{eqnarray}\label{SD}
\Phi_N(\kappa)=AC_N^2\kappa^{-11/3}\exp\left(-\kappa^2/k_{\mathrm{m}}^2\right),
\end{eqnarray}
where the refractive index structure function $C_N^2\simeq C_n^2/n^2$, the constant $A\approx 0.033$, the minimum wave-number $k_{\mathrm{m}}=2\pi/L$, and the above form is valid for $\kappa\gg2\pi/L$~\citep{2002Spangler,2023Kontar}. After some calculations, we once again obtain the wavelength-independent estimation of GDP (model-independent: not limited to plasma emission mechanism) as
\begin{eqnarray}\label{ep2}
\langle e^{\prime 2}\rangle\equiv
{2\pi^4L^2}\left[\int_0^{\infty}\kappa^3\Phi_N(\kappa)\mathrm{d}\kappa\right]^2
=\mathcal{N}_1\frac{C_N^4L^2}{k_{\mathrm{m}}^{-2/3}}
\end{eqnarray}
and DDP that is proportional to the distance $L$ as
\begin{eqnarray}\label{epp2}
\langle e^{\prime\prime 2}\rangle\equiv\frac{\pi^2 L}{2k^2}\int_0^{\infty}\kappa^5\Phi_N(\kappa)\mathrm{d}\kappa
=\mathcal{N}_2\frac{C_N^2L}{k^2k_{\mathrm{m}}^{-7/3}},
\end{eqnarray}
where the coefficients
$\mathcal{N}_1={\pi^4}A^2\Gamma^2(1/6)/2\approx1.643$ and $\mathcal{N}_2={\pi^2}A\Gamma(7/6)/4\approx0.076$. The results are depicted in the lower panels of Figure \ref{fig:Figure3}. We introduce the relative depolarization ratio $\eta$ to measure the impact of GDP and DDP on radio waves which is given by
\begin{eqnarray}\label{eta}
\eta\equiv\frac{\sqrt{\langle e^{\prime 2}\rangle}}{\sqrt{\langle e^{\prime\prime 2}\rangle}}.
\end{eqnarray}
In our case, the ratio $\eta\simeq10^4-10^5\gg1$, which means that the impact of GDP on polarized radio waves is not only non-negligible but also significantly greater than the impact of DDP. For simplicity, we can perform a quick calculation to confirm the results we reached in the upper panels of the Figure~\ref{fig:Figure3}, let $C_N^2=10^{-11}~\textrm{cm}^{-2/3}$ (strong scattering case), $k_{\textrm{m}}=10^{-5}~\textrm{cm}^{-1}$, and $\lambda=300~\textrm{cm}$, using Equations~(\ref{ep2}) and (\ref{epp2}), then we have $\langle e^{\prime 2}\rangle\approx4.13$ and $\eta\approx17538\gg1$.

\section{Discussion and Conclusions} \label{sec:Discussion and Conclusions}
\subsection{Discussion} \label{Discussion}

Our aforementioned calculations are not only applicable to the propagation of linearly polarized radio waves in the turbulent plasma of the solar wind but also to more general cases, such as the radio emission from the magnetosphere of pulsars, which is used to explain the phenomenon of fast radio bursts. According to \citet{1938Rytov}, the Rytov rotation happens, as the radio wave propagates along a non-planar curve. The GDP effect examined in isotropic and smoothly inhomogeneous scattering of linearly polarized radio waves results from the superposition of randomly oriented polarizations of the waves traveling along different random paths.

However, the situation is different for circular polarization \citep{2014Gorodnichev}. Circularly polarized light can be considered as a superposition of two linearly cross-polarized radio waves shifted in phase by $\pi/2$. In the case of isotropic coronal scattering, both components of the polarized waves will undergo Rytov rotation, but the phase shift between them remains unchanged. Therefore, a circularly polarized radio wave traveling along any random trajectory is unaffected by the Rytov rotation.

\begin{figure*}[t!]
  \centering
  \includegraphics[width=1.0\linewidth]{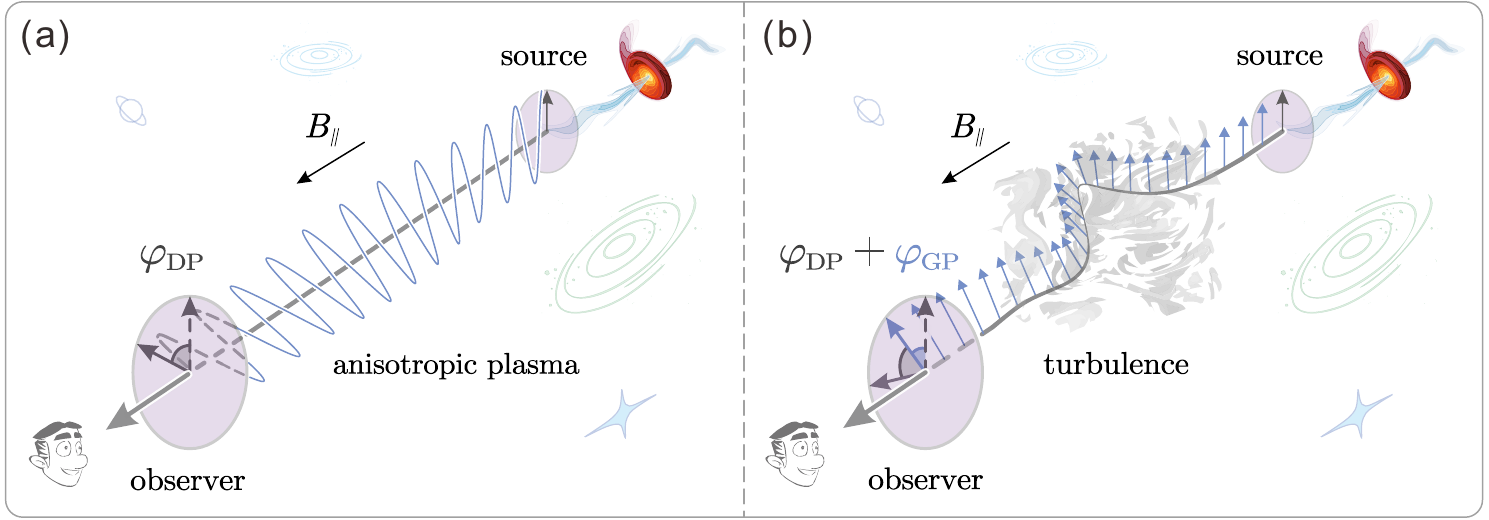}
  \caption{Cartoon illustration of depolarization effects. (a) Ideal case: the propagation of radio waves in an anisotropic plasma with uniform distribution along the line-of-sight. The dynamical phase $\varphi_{DP}$ caused by Faraday rotation effect. (b) Actual case: total phase of depolarization in a turbulent magnetized plasma equals to the dynamical phase $\varphi_{DP}$ caused by Faraday rotation effect plus the geometric phase $\varphi_{GP}$ (blue arrows represent the nontrivial rotation to the polarization plane) caused by the twisting path with non-zero torsion (the diagram has, to some extent, magnified the degree of distortion of the path of the ray.).
  \label{fig:Figure5}}
\end{figure*}

It should be noted that the focus of polarimetric solar radio studies has been solely on circular polarization for decades \citep{2022Kansabanik}. \citet{1973Grognard} concluded that it is hard to detect any linear polarization in solar radio emissions at metre-wavelength (30-300 MHz) due to the large differential Faraday rotation experienced by the emission while traveling through the corona will suffer depolarization. Chapter 6 of \citet{2023Kansabanik} discussed the reasons in detail. However, \citet{2022Dey} found that some type III bursts show the presence of linearly polarized emission. Recently, \citet{2024Dey} presented the first robust imaging-based evidence for linearly polarized emission in solar radio bursts at metre-wavelength. Therefore, our analysis still remains valid in solar radio bursts.
In general, there are countless discussions on the depolarization of electromagnetic waves by turbulent media. However, we have discussed the GDP phenomenon induced by the propagation of linearly polarized radio waves in astrophysical turbulent plasma from a new perspective. The further refinement of this method still requires development from the following aspects:
\begin{itemize}
    \item Ray tracing techniques \citep{2019Kontar,2020Chen}.
Numerical ray tracing simulations of radio waves propagating through plasma resulted in the polarization position angles displaying wavelength dependencies \citep{2024Lower}. The impact of wavelength dependence on Rytov rotation should be considered in other plasma environments. Detailed numerical simulation can provide credible information on the curvature and torsion of the ray paths, then the geometric phase can be calculated.
\item Faraday conversion and rotation effects. As we mentioned before, Rytov rotation is an example of the geometric phase. According to \citet{2004Bliokh}, the geometrical optics approximation in an isotropic and smoothly inhomogeneous medium is anisotropic. \citet{1976Budden} has shown that the additional `phase memory' (one type of geometric phase) is important for radio waves in an isotropic ionosphere. From \citet{1986Berry} and \citet{1990Berry}, we can conclude that if the plasma is anisotropic, with an axis of birefringence, and an axis of gyrotropy that is locally fixed by the direction of the background magnetic field, then the slow rotation of these axes will produce a geometric phase. In the radio frequency band, the Faraday rotation effect will cause any linearly polarized wave to lose its initial polarization orientation characteristics. \citet{2012Liu} discussed the impact of the geometric phase on Faraday rotation and concluded that the magnitudes of geometric and dynamical Faraday rotation angles are of the same order using typical parameters of the Tokamak plasma. It might be used for astrophysical plasma diagnostics, as shown in Figure~\ref{fig:Figure5}. Therefore, a reconsideration of the path-dependence depolarization in the Faraday rotation measure could refresh our understanding of the strength and geometry of astrophysical magnetic fields. The interplay between works in the community of optics and those in the community of astrophysics, such as \citealt{2000Macquart} and \citealt{2022Lyutikov} may inspire more new ideas for both.
\item Combining with more observations. The GDP effect deserves the attention of radio astronomers. With the rapid development of radio observation technology, such as the Square Kilometre Array (SKA), the Five-hundred-meter Aperture Spherical Telescope (FAST), and the Very Large Array (VLA) et al., we are eager to gain a deeper understanding of the radio sources (pulsar magnetospheres, blazar jets, gamma-ray bursts afterglows, fast radio bursts (FRBs) and other radio transient phenomena) via polarization information of radio waves \citep{1981Altunin,2019Urata,2020Luo,2022Kansabanik,2023Kansabanik,2023Kumar,2023Zhang,2024Lower}. Recently, \citet{2022Bastian} reported detections of linear polarization in stellar radio bursts. \citet{2022Feng} reported the detection of highly linearly polarized radio emissions from the fast radio burst sources, such as FRB 20121102A, FRB 20180916B, and FRB 20201124A show frequency-dependent behaviors. Hence, the consideration of the frequency dependence of the GDP effect for the propagation of radio waves, as well as the discussion on the robustness of the geometric phase against ambient perturbations, are worth further research. Additionally, it is worth mentioning that non-stationary gravitational lensing and plasma lensing effects which could bend the path may also affect the polarization vector of electromagnetic waves \citep{1992Dyer,2018Crisnejo,2023Er}.
\end{itemize}

\subsection{Conclusions} \label{Conclusions}

To sum up, we demonstrate an important but long-neglected decoherence mechanism in astrophysics. We propose a toy model which including the three-dimensional stochastic description of rays and the techniques of differential geometry to estimate the impact of GDP on the angle-of-arrival of polarized radio waves to the Earth's observer. Based on the geometrical optics approximation, we assume that the radio waves generated in the outer solar corona during flares are created through the mechanisms of harmonic emission of plasma for simplicity. Thus, our results become independent of the wavelength. Then we applied the three-dimensional stochastic description of ray and the techniques of differential geometry for analysis. In the case of isotropic coronal scattering, the angle-of-arrival of the linearly polarized radio waves duo to GDP can not be neglected (in the isotropic case). This is the first time the concept and related methods of geometric phase have been introduced into the field of radio astronomy to study polarization issues. Our results can be generalized to other astrophysical radio burst phenomena that exhibit coherent radiation characteristics, such as radio emission from pulsar magnetospheres triggered by plasma instabilities (in the anisotropic case). Our results are expected to profoundly influence the analysis of astrophysical depolarization, and it is essential for understanding astrophysical processes through a substantial amount of polarization information from electromagnetic radiation.

\vspace{5mm}
\noindent{Ze-Lin Zhang would like to thank Xiang-Yu Wang, Jian Liu, and Pei-Jin Zhang  for useful discussions. This work is supported by the Natural Science Foundation of Education Department of Anhui Province under Grant No. 2024AH050330 and the Talents Introduction Project of Anhui Science and Technology University under Grant No.~DQYJ202202 and the National Natural Science Foundation of China under the grant No.~12393852.}

\vspace{5mm}
%% Similar to \facility{}, there is the optional \software command to allow
%% authors a place to specify which programs were used during the creation of
%% the manuscript. Authors should list each code and include either a
%% citation or url to the code inside ()s when available.

%% Include this line if you are using the \added, \replaced, \deleted
%% commands to see a summary list of all changes at the end of the article.
%\listofchanges

\bibliography{sample631}{}
\bibliographystyle{aasjournal}
\end{document}